\documentclass[times,authoryear]{elsarticle}

\usepackage{jasr}
\usepackage{framed,multirow}

\usepackage{amsmath}
\usepackage{amssymb}
\usepackage{latexsym}

\usepackage{url}
\usepackage[dvipsnames]{xcolor}
\definecolor{newcolor}{rgb}{.8,.349,.1}

\usepackage[citebordercolor=white]{hyperref}

\journal{Advances in Space Research}

\begin{document}
\newcommand{\glm}[1]{\textcolor{black}{#1}}
\newcommand{\mai}[1]{\textcolor{black}{#1}}
\newcommand{\others}{\textit{et al.}}
\newcommand{\aap}{\textit{Astronomy \& Astrophysics}}
\newcommand{\aapr}{\textit{Astronomy \& Astrophysics Reviews}}
\newcommand{\ao}{\textit{Applied Optics}}
\newcommand{\apj}{\textit{The Astrophysical Journal}}
\newcommand{\apjs}{\textit{The Astrophysical Journal Supplement}}
\newcommand{\jqsrt}{\textit{Journal of Quantitative Spectroscopy \& Radiative Transfer}}
\newcommand{\mnras}{\textit{Monthly Notices of the Royal Astronomical Society}}
\newcommand{\nat}{\textit{Nature}}
\newcommand{\prd}{\textit{Physics Review D}}
\newcommand{\bmat}[1]{\mathrm{\textbf{#1}}}
\newcommand{\bvec}[1]{\textit{\textbf{#1}}}
\newcommand{\de}{\mathrm{d}}

\verso{Giovanni La Mura \textit{et al.}}

\begin{frontmatter}

\title{Interstellar dust as a dynamic environment}%

\author[1]{Giovanni \snm{La Mura}\corref{cor1}}
\cortext[cor1]{Corresponding author: 
  Tel.: +39-070-71180-234
}
\ead{giovanni.lamura@inaf.it}
\author[1]{Giacomo \snm{Mulas}}
\ead{giacomo.mulas@inaf.it}
\author[2]{Maria Antonia \snm{Iat\`{i}}}
\ead{mariaantonia.iati@cnr.it}
\author[3]{Cesare \snm{Cecchi-Pestellini}}
\ead{cesare.cecchipestellini@inaf.it}
\author[4,2,5]{\\Shadi \snm{Rezaei}}
\ead{shadi.rezaei@unime.it}
\author[4,2]{Rosalba \snm{Saija}}
\ead{rosalba.saija@unime.it}

\affiliation[1]{organization={INAF - Osservatorio Astronomico di Cagliari},
                addressline={Via della Scienza 5},
                city={Selargius},
                postcode={09047},
                country={Italy}}

\affiliation[2]{organization={CNR - Istituto per i Processi Chimico-Fisici},
                addressline={Viale F. Stagno d'Alcontres 37},
                city={Messina},
                postcode={98158},
                country={Italy}}

\affiliation[3]{organization={INAF - Osservatorio Astronomico di Palermo},
                addressline={Piazza del Parlamento 1},
                city={Palermo},
                postcode={90134},
                country={Italy}}

\affiliation[4]{organization={Universit\`a di Messina - Dip. di Scienze Matematiche e Informatiche, Scienze Fisiche e Scienze della Terra},
                addressline={Viale F. Stagno D'Alcontres 31},
                city={Messina},
                postcode={98166},
                country={Italy}}
                
\affiliation[5]{organization={University of Kurdistan,Dept. of Physics, Faculty of Science},
                addressline={Pasdaran Street},
                city={Sanandaj},
                postcode={416/6613566176},
                country={Iran}}

\received{14 January 2025}
\finalform{29 April 2025}
\accepted{1 May 2025}
\availableonline{7 May 2025}
\communicated{S. Sarkar}


\begin{abstract}
  In spite of accounting for only a small fraction of the mass of the Interstellar Medium (ISM), dust plays a primary role in many physical and chemical processes in the Universe. It is the main driver of extinction of radiation in the UV/optical wavelength range and a primary source of thermal IR emission. Dust grains contain most of the refractory elements of the ISM and they host chemical processes that involve complex molecular compounds. However, observational evidence suggests that grain structure is highly non-trivial and that dust particles are characterized by granularity, asymmetry and stratification, which significantly affect their interaction with radiation fields. Accurate modeling of such interaction is fundamental to properly explain observational results, but it is a computationally demanding task. Here we present the possibility to investigate the effects of radiation/particle interactions in non-spherically symmetric conditions using a novel implementation of the Transition Matrix formalism, designed to run on scalable parallel hardware facilities.
\end{abstract}


\end{frontmatter}


\section{Introduction}\label{sec_intro}

Dust is a fundamental component of the Interstellar Medium (ISM). The most striking evidence of its presence can be observed through the extinction effects that it produces in dense environments, such as molecular clouds, star-forming regions, and in the central regions of active galactic nuclei (AGN). In the Milky Way, dust represents approximately one percent of the mass of the ISM, but it is the dominant form of heavy refractory elements. There is ample agreement that interstellar dust particles are composed mainly of silicates and carbonaceous materials \glm{\citep{Aiello00, Mulas13, Mishra15}}, with the addition of volatile ices enriched with complex molecular compounds. The way in which these ingredients compose actual dust grains, whether silicates or carbonaceous materials are in completely separate dust particles \glm{\citep[e.g.][]{Weingartner01}} or are intermixed \glm{\citep{Li97, Hensley23}}, whether several dust size populations are present, as well as how complex are the shapes of these particles, is still poorly constrained. The effects of dust are observed all the way up to the high redshift Universe, where they trace the evolution and the chemical history of galaxies \glm{\citep{Zavala21, Witstok23, Yang23}}. However, exposure to various types of astrophysical environments necessarily leads to differences in the structure and composition of dust particles \glm{\citep{Cecchi-Pestellini14}}.

The interaction between dust and radiation is not limited to the effect of extinction. As particles scatter, absorb, and emit photons, a wealth of dynamical, chemical, and thermodynamic processes take place. The balance between gravitational and radiation pressure effects may lead dust to migrate away from the plane of disk galaxies and even to be blown out of star-forming galaxies into the inter-galactic medium \glm{\citep{Greenberg87, Alton94, Alton99}}. Moreover, non-symmetric particles experience radiative torques that can cause them to spin up so efficiently to fragment due to centrifugal stress \glm{\citep{Li97, Hoang21}} \mai{with the efficiency of such processes depending critically on the radiation extinction cross-section of the particles.} This parameter is not easy to constrain, as its evaluation out of spherically symmetric cases is a computationally demanding task. 
\mai{The Transition matrix (T-matrix) formalism  \citep{Waterman71} offers a valuable tool for an accurate calculation of dust extinction cross section and of all the quantities characterizing dust optical and opto-mechanical behaviour. Many implementations of the formalism are available in the literature \citep[e.g.][]{Barber90, Saija01, Borghese07, Mackowski13}. \citet{Saija01}, in particular, used the T-matrix approach to} investigate the dependence of the extinction cross-sections on the structure of dust particles and on their composition. They found that porosity and non-spherical symmetry have a large effect on the interactions between radiation and dust particles. In this paper, we present the first results obtained with a new implementation of the T-matrix method \glm{applied by \citet{Saija01}} that, taking advantage of modern computing technologies, such as GPU acceleration and multi-core processing facilities, reduces the computing times required to solve complex, realistic particle models.

This paper is structured as follows: in section~\ref{sec_theory} we summarize the theoretical framework, based on the T-matrix, to solve the scattering problem; in section~\ref{sec_model} we describe our particle model; in section~\ref{sec_results} our results are presented and discussed, and, finally, our conclusions are summarized in section~\ref{sec_conclusions}.

\section{Theoretical framework}\label{sec_theory}

The interaction of material particles with electromagnetic radiation may occur in different regimes, controlled by the size parameter:
\begin{equation}
  x = \frac{2 \pi n_m}{\lambda} \rho,
\end{equation}
where $n_m$ is the refractive index of the medium where the particle is embedded and $\rho$ is the radius of the smallest sphere containing the particle. \glm{In astrophysical environments, the medium is often emptier than the best vacuum that we can create in laboratory. Therefore, for our purposes, we can adopt $n_m = 1$ in $x$. Calling $n$ the refractive index of the particle's material, if $x\, |n| \ll 1$ (i.e. when the optical path of radiation in} the particle is significantly smaller than the incident radiation wavelength),\footnote{\glm{For conductive materials such as Fe, Mg and graphite, $n$ increases with wavelength, so that we can have $x\, |n| \geq 1$ even if $x \ll 1$ \citep{Li03}.}} the interaction occurs under Rayleigh approximation \glm{\citep{Bohren83}}, which implies a scattering process having cross-section $\sigma_{SCA} \propto \lambda^{-4}$. If, on the other hand, we have $x\, |n| \gg 1$, the problem can be addressed in the framework of ray optics. Conversely, whenever $x\, |n| \sim 1$ (typically in the $0.1 \leq x\, |n| \leq 10$ range) a full treatment of the problem, with a proper solution of Maxwell's field equations, is required. Such solution exists, in the case of plane waves interacting with spherically symmetric particles, in the framework of the \textit{Mie theory} \citep{Mie1908}. More general cases, instead, have no exact solution and need to be addressed with numerical approaches.

Let's start from considering the case of a mono-chromatic radiation field, with electric vector component of the form:
\begin{equation}
    \bvec{E}(\bvec{r}, t) = \mathfrak{R} \left[ \bvec{E}(\bvec{r}, t) \exp(-i \omega t) \right]
\end{equation}
interacting with a particle. The electromagnetic field obeys the Helmholtz equations:
\begin{subequations}
\begin{align}
  (\nabla^2 + n^2 k_v^2)\, \bvec{E} & = 0 \label{elec_Helm} \\
  (\nabla^2 + n^2 k_v^2)\, \bvec{B} & = 0  \label{mag_Helm}
\end{align}
\end{subequations}
with the conditions:
\begin{equation}
  \begin{array}{lcl}
    \nabla \cdot \bvec{E} = 0 & \& & \qquad \nabla \cdot \bvec{B} = 0 \\
    \\
    \nabla \times \bvec{E} = i k_v \bvec{B} & \& & \nabla \times \bvec{B} = -i k_v n^2 \bvec{E}, \\
  \end{array} \nonumber
\end{equation}
where we introduced the magnitude of the propagation vector in vacuum $k_v = \omega / c$ and the complex material refractive index:
\begin{equation}
  n = \sqrt{\mu \left( \varepsilon + \frac{4 \pi i \sigma}{\omega} \right)},
\end{equation}
with $\mu$ being the magnetic permeability, $\varepsilon$ the dielectric constant and $\sigma$ the conductivity of the material. Here, we assume that the incident field is a linearly polarized wave of the form:
\begin{equation}
    \bvec{E}_I = E_0 \hat{\bvec{e}}_I \exp(i \bvec{k}_I \cdot \bvec{r}), \label{eq_def_pol_field}
\end{equation}
where:
\begin{equation}
    \bvec{k}_I = n k_v \hat{\bvec{k}}_I \quad \mathrm{and} \quad \hat{\bvec{k}}_I \cdot \hat{\bvec{e}}_I = 0 \nonumber
\end{equation}
and we omit the dependence on time for convenience. Since the scattered field $\bvec{E}_S$ needs to obey the Helmholtz equation Eq.~(\ref{elec_Helm}), calling $\psi$ any of its rectangular components, we can write:
\begin{equation}
  (\nabla^2 + n^2 k_v^2) \psi(\bvec{r}, k) = 0. \label{eq_helm_comp}
\end{equation}
The solutions of Eq.~(\ref{eq_helm_comp}) that describe an actual scattering process must satisfy boundary conditions on the surface of the particle and are, in general, dependent on the incidence direction $\hat{\bvec{k}}_I$ and the observation direction $\hat{\bvec{k}}_S = \hat{\bvec{r}}$. Choosing a reference frame with origin in the particle, the solution of Eq.~(\ref{eq_helm_comp}) can be expanded in a series of \textit{spherical harmonics} \citep{Borghese07}:
\begin{equation}
  \psi(\bvec{r}, k) = \sum_{lm} C_{lm}(\hat{\bvec{k}}_I) Y_{lm}(\hat{\bvec{r}}) R_l(r, k) / r, \label{eq_har_exp}
\end{equation}
where the amplitudes $C_{lm}$ are determined by the boundary conditions and the spherical harmonics are:
\begin{equation}
  Y_{lm}(\hat{\bvec{r}}) \equiv Y_{lm}(\theta, \phi) = \sqrt{\frac{2 l + 1}{4 \pi} \frac{(l - m)!}{(l + m)!}} P_{lm}(\cos \theta) \mathrm{e}^{i m \phi},
\end{equation}
with $P_{lm}$ representing the associated Legendre functions of the first kind. Using Eq.~(\ref{eq_har_exp}) in Eq.~(\ref{eq_helm_comp}) leads to an expression of the radial terms:
\begin{equation}
  \frac{\de^2 R_l}{\de^2 r} + \left[ k^2 - \frac{l (l + 1)}{r^2}\right] R_l = 0, \nonumber
\end{equation}
which, in the \textit{far zone} approximation (i.e. in the assumption of large $r$), simplifies to:
\begin{equation}
  \frac{\de^2 R_l}{\de^2 r} + k^2 R_l = 0. \label{eq_radial_sol}
\end{equation}
Since the scattered wave needs to satisfy the \textit{radiation condition}, implying that it must assume the form of an outgoing spherical wave at large distances \citep{Stratton41,Ishimaru78}, the solutions of Eq.~(\ref{eq_radial_sol}) are of the form:
\begin{equation}
  R_l = E_0 \exp(i k r) \label{eq_radial_form}
\end{equation}
independently of $l$. Defining the components of the \textit{normalized scattering amplitude} as:
\begin{equation}
  f_\alpha(\hat{\bvec{k}}_S, \hat{\bvec{k}}_I) = \sum_{lm} C_{lm}(\hat{\bvec{k}}_I) Y_{lm}(\hat{\bvec{r}}), \label{eq_norm_ampl}
\end{equation}
and taking into account Eq.~(\ref{eq_radial_form}), Eq.~(\ref{eq_har_exp}) becomes:
\begin{equation}
  \psi_\alpha(\bvec{r}, k) = E_0 f_\alpha(\hat{\bvec{k}}_S, \hat{\bvec{k}}_I) \frac{\exp(i k r)}{r},
\end{equation}
where $\alpha$ is the component index, so that the scattered radiation field can be asymptotically expressed in vector form as:
\begin{equation}
  \bvec{E}_S = E_0 \frac{\exp(i k r)}{r} \bvec{f}(\hat{\bvec{k}}_S, \hat{\bvec{k}}_I). \label{eq_asympt_sca}
\end{equation}

For general particles, the field described by Eq.~(\ref{eq_asympt_sca}) depends on the directions of incident and scattered radiation, as well as on the orientation of the particle. In the spherically symmetric case, on the contrary, the dependence is limited only to the angle between the incident and the scattered radiation, which we call the \textit{scattering angle}. In addition, using the constraint $\nabla \cdot \bvec{E} = 0$, we can derive:
\begin{equation}
  \nabla \cdot \bvec{E}_S = E_0 \left( \nabla \frac{\exp(i k r)}{r} \right) \cdot \bvec{f} + E_0 \frac{\exp(i k r)}{r} \nabla \cdot \bvec{f} = 0,
\end{equation}
from which we infer that the condition $\hat{\bvec{k}}_S \cdot \bvec{f} = 0$ also implies $\nabla \cdot \bvec{f} = 0$.

The scattering amplitude is related to the amount of electromagnetic energy flux that the particle scatters per unit solid angle in any given direction and it provides a full description of the scattering process. The incoming flux of electromagnetic energy is given by the Poynting vector:
\begin{equation}
  \langle \bvec{S}_I \rangle = \frac{c}{8 \pi} \mathfrak{R} (\bvec{E}_I \times \bvec{B}_I^*) = \frac{n c}{8 \pi} |E_0|^2 \hat{\bvec{k}}_I,
\end{equation}
where the $^*$ symbol denotes complex conjugation. Similarly, the scattered energy far from the particle can be expressed as:
\begin{equation}
  \langle \bvec{S}_S \rangle = \frac{c}{8 \pi} \mathfrak{R} (\bvec{E}_S \times \bvec{B}_S^*) = \frac{n c}{8 \pi} |\bvec{E}_S|^2 \hat{\bvec{k}}_S.
\end{equation}
We define the \textit{differential scattering cross-section} as the ratio of the energy scattered in a given direction $\hat{\bvec{k}}_S$ with respect to the incoming energy flux:
\begin{equation}
  \frac{\de \sigma_{SCA}}{\de \Omega} = \lim_{r \rightarrow \infty} \frac{r^2 |\langle \bvec{S}_S \rangle|}{|\langle \bvec{S}_I \rangle|} = |\bvec{f}(\hat{\bvec{k}}_S, \hat{\bvec{k}}_I)|^2, \label{eq_diff_cs}
\end{equation}
so that the ratio of the total amount of energy flux scattered by the particle with respect to the incoming one is:
\begin{equation}
  \sigma_{SCA} = \int_\Omega \frac{\de \sigma_{SCA}}{\de \Omega}\, \de \Omega = \int_\Omega |\bvec{f}(\hat{\bvec{k}}_S = \hat{\bvec{r}}, \hat{\bvec{k}}_I)|^2\, \de \Omega.
\end{equation}

In addition to $\sigma_{SCA}$, we can also define the \textit{absorption cross-section} $\sigma_{ABS}$ as the ratio among the amount of energy flux that is absorbed by the particle and the incoming flux:
\begin{equation}
  \sigma_{ABS} = -\frac{1}{|\bvec{S}_I|} \frac{c}{8 \pi} \int \mathfrak{R} (\bvec{E} \times \bvec{B}^*) \cdot \hat{\bvec{n}} \de S, \label{eq_def_abs}
\end{equation}
where $\hat{\bvec{n}}$ is the unit outward vector of the particle's surface. Using the Gauss theorem, Eq.~(\ref{eq_def_abs}) becomes:
\begin{equation}
  \sigma_{ABS} = -\frac{1}{|\bvec{S}_I|} \frac{c}{8 \pi} \int_V \mathfrak{R} [\nabla \cdot (\bvec{E} \times \bvec{B}^*)] \de V,
\end{equation}
which, by means of the mixed vector product identity and with the assumed form of the electric and magnetic fields, eventually becomes:
\begin{equation}
  \sigma_{ABS} = \frac{1}{|\bvec{S}_I|} \frac{c}{8 \pi} \int_V \mathfrak{R} (i k_v {n^*}^2 |\bvec{E}|^2 - i k_v |\bvec{B}|^2) \de V. \label{eq_res_abs}
\end{equation}
Eq.~(\ref{eq_res_abs}) shows that absorption arises in materials with complex refractive index $n$.

In the end, we are able to evaluate the \textit{total extinction} cross-section, as the sum of the contributions from scattering and absorption:
\begin{equation}
   \sigma_{EXT} = \sigma_{SCA} + \sigma_{ABS}.
\end{equation}
Since these cross-sections are, in general, different from the geometric cross-section of the particle $G$, as seen from the direction of the incident radiation, it is customary to introduce the scattering, absorption and extinction efficiencies, defined as:
\begin{equation}
  Q_{SCA} = \frac{\sigma_{SCA}}{G}; \quad Q_{ABS} = \frac{\sigma_{ABS}}{G}; \quad Q_{EXT} = \frac{\sigma_{EXT}}{G}.
\end{equation}

\begin{figure}
\begin{center}
\includegraphics[width=0.9\textwidth]{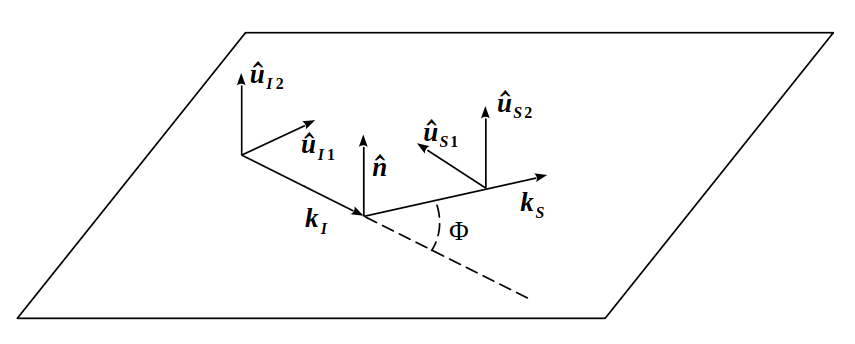}
\end{center}
\caption{The scattering plane. $\Phi$ is the scattering angle, $\bvec{k}_I$ is the incident wave vector, $\bvec{k}_S$ is the scattered wave vector, $\hat{\bvec{n}}$ is the plane normal unit vector, while $\hat{\bvec{u}}_{I \eta}$ and $\hat{\bvec{u}}_{S \eta}$ ($\eta = 1,2$) are the unit base vectors for the scattering process description. Adapted from \citet{Borghese07}. \label{fig_sca_plane}}
\end{figure}
\subsection{The T-matrix}

When the particle has no spherical symmetry, the relation between the incident and the scattered fields cannot be derived analytically and the problem needs to be solved through numerical approaches. However, an analytical solution can be recovered if the particle is constituted (or can be approximated) by an aggregate of spherical monomers. Following the approach described by \citet{Borghese79} and further detailed by \citet{Saija01}, it is possible to represent a particle of arbitrary shape as an aggregate of spherical monomers and to use the linearity of the field equations, which describe the field interaction with each monomer, to derive the total solution. Assuming again that the incident field is a polarized plane wave as in Eq.~(\ref{eq_def_pol_field}), it is useful to express the polarization of the field with respect to the scattering plane, defined by the direction of incidence $\hat{\bvec{k}}_I$ and of scattering $\hat{\bvec{k}}_S$. We can then define two pairs of mutually orthogonal unit vectors $\hat{\bvec{u}}_{I,\eta}$ and $\hat{\bvec{u}}_{S,\eta}$ ($\eta = 1,2$), illustrated inf Fig.~\ref{fig_sca_plane}, chosen in such a way that $\hat{\bvec{u}}_{I,1}$ and $\hat{\bvec{u}}_{S,1}$ lie in the scattering plane, $\hat{\bvec{u}}_{I,2}$ and $\hat{\bvec{u}}_{S,2}$ are orthogonal to it, and:
\begin{equation}
  \begin{array}{l}
    \hat{\bvec{u}}_{I,1} \times \hat{\bvec{u}}_{I,2} = \hat{\bvec{k}}_I \\
    \\
    \hat{\bvec{u}}_{S,1} \times \hat{\bvec{u}}_{S,2} = \hat{\bvec{k}}_S. \\
  \end{array} \nonumber
\end{equation}
Then, the incident and the scattered fields can be decomposed as:
\begin{equation}
  \begin{array}{l}
    \bvec{E}_I = \sum_\eta (\bvec{E}_I \cdot \hat{\bvec{u}}_{I \eta}) \hat{\bvec{u}}_{I \eta} = \sum_\eta \bvec{E}_{I\eta} \\
    \\
    \bvec{E}_{S} = \sum_{\eta} (\bvec{E}_{S} \cdot \hat{\bvec{u}}_{S\eta}) \hat{\bvec{u}}_{S\eta} = \sum_\eta \bvec{E}_{S\eta} \\
  \end{array} \nonumber
\end{equation}

Following \citet{Borghese07}, we can expand the incident electric field as:
\begin{equation}
  \bvec{E}_{I\eta}(\bvec{r}) = E_{0\eta} \sum_{plm} \bvec{J}^{(p)}_{lm}(\bvec{r}, k) W^{(p)}_{I\eta lm}, \label{eq_inc_exp}
\end{equation}
where $\bvec{J}^{(p)}_{nlm}$ are vector multipole fields and the amplitudes of the incident field are $W^{(p)}_{I\eta lm} = W^{(p)}_{lm}(\hat{\bvec{u}}_{I\eta}, \hat{\bvec{k}}_I)$, with:
\begin{equation}
  W^{(p)}_{lm}(\hat{\bvec{e}}, \hat{\bvec{k}}) = 4\pi i^{p + l - 1} \hat{\bvec{e}} \cdot \bvec{Z}^{(p)*}_{lm}(\hat{\bvec{k}}). \label{eq_transverse}
\end{equation}
These amplitudes are such that the result of the linear combination of $\bvec{J}_{lm}^{(p)}$ in Eq.~(\ref{eq_inc_exp}) is aligned with $\hat{\bvec{u}}_{I\eta}$. In Eq.~(\ref{eq_transverse}), $\bvec{Z}^{(p)*}_{lm}(\hat{\bvec{k}})$ are transverse harmonics \citep{Fucile97, Saija03b}. Similarly, the scattered field can be expressed as:
\begin{equation}
  \bvec{E}_{S\eta}(\bvec{r}) = E_{0\eta} \sum_{plm} \bvec{H}^{(p)}_{lm}(\bvec{r}, k) A^{(p)}_{\eta lm},
\end{equation}
where $A^{(p)}_{\eta lm}$ are the amplitudes of the scattered field. Thanks to the linearity of the Maxwell equations and forcing continuity on the surface of the particle, such amplitudes can be expressed as a function of the incident field's amplitudes introducing the T-matrix as \citep{Waterman71}:
\begin{equation}
  A^{(p)}_{\eta lm} = S^{(pp')}_{lml'm'} W^{(p')}_{I\eta l' m'}. \label{eq_intro_tm}
\end{equation}
The elements of the T-matrix, $S^{(pp')}_{lml'm'}$, contain all the information concerning the particle morphology and its orientation with respect to the incident radiation field. The existence of such a T-matrix connecting the incident and scattering amplitudes implies no assumption on the particle, as it just stems from Maxwell's equations and their linearity. For spherical particles, the T-matrix approach completely overlaps with the Mie solution of the light scattering problem. 

Let's assume that the particle interacting with the incident wave of Eq.~(\ref{eq_def_pol_field}) can be modeled as an aggregate of spheres, identified by an index $\alpha$, having radius $\rho_\alpha$, center coordinates $\bvec{R}_\alpha$, refractive index $n_\alpha$ and embedded in a non-absorbing medium with refractive index $n$.  The incident field can still be expanded as in Eq.~(\ref{eq_inc_exp}), while the scattered field takes the form:
\begin{equation}
  \bvec{E}_{S\eta} = E_{0 \eta} \sum_\alpha \sum_{plm} \bvec{H}^{(p)}_{lm}(\bvec{r}_\alpha, k) \mathcal{A}^{(p)}_{\eta\alpha lm}, \label{eq_sca_exp}
\end{equation}
where $\bvec{r}_\alpha = \bvec{r} - \bvec{R}_\alpha$. Conversely, the field inside the $\alpha$-th sphere expands to:
\begin{equation}
  \bvec{E}_{T\eta\alpha} = E_{0 \eta} \sum_\alpha \sum_{plm} \bvec{J}^{(p)}_{lm}(\bvec{r}_\alpha, k_\alpha) \mathcal{C}^{(p)}_{\eta\alpha lm}, \label{eq_inter_exp}
\end{equation}
with $k_\alpha = n_\alpha k$. The amplitudes $\mathcal{A}^{(p)}_{\eta\alpha lm}$ and $\mathcal{C}^{(p)}_{\eta\alpha lm}$ in Eqs.(\ref{eq_sca_exp}) and (\ref{eq_inter_exp}) are determined by the boundary conditions at the surface of each sphere and their knowledge solves the problem of scattering by the whole aggregate. In fact, the amplitudes $\mathcal{A}^{(p)}_{\eta\alpha lm}$ are the solution of the linear system of non-homogeneous equations \citep{Borghese84}:
\begin{equation}
  \sum_{\alpha'} \sum_{p' l' m'} \mathcal{M}^{(pp')}_{\alpha l m \alpha' l' m'} \mathcal{A}^{(p')}_{\eta \alpha' l' m'} = -\mathcal{W}^{(p)}_{\eta \alpha l m}, \label{eq_lin_system}
\end{equation}
where we have introduced the \textit{shifted} amplitudes of the incident field:
\begin{equation}
  \mathcal{W}^{(p)}_{\eta \alpha l m} = \sum_{p' l' m'} \mathcal{J}^{(pp')}_{\alpha l m 0 l' m'} W^{(p')}_{I\eta l' m'}, \label{eq_shifted_ampl}
\end{equation}
and:
\begin{equation}
  \mathcal{M}^{(p p')}_{\alpha l m \alpha' l' m'} = \left[ R^{(p)}_{\alpha l} \right]^{-1} \delta_{\alpha \alpha'} \delta_{p p'} \delta_{l l'} \delta_{m m'} + \mathcal{H}^{(p p')}_{\alpha l m \alpha' l' m'}. \label{eq_deltas}
\end{equation}
The quantities $\mathcal{J}^{(pp')}_{\alpha l m 0 l' m'}$ are the elements of the matrix that translates the multipole fields from the origin of the coordinates to the center of the $\alpha$-th sphere $\bvec{R}_\alpha$, according to the multipole field addition theorem \citep{Borghese80}. The quantities $R^{(1)}_{\alpha l}$ and $R^{(2)}_{\alpha l}$, except for a sign, are the elements of the T-matrix of the $\alpha$-th sphere, while the quantities $\mathcal{H}^{(p p')}_{\alpha l m \alpha' l' m'}$ are also derived from the vector multipole field addition theorem and take into account the effects of multiple scattering processes between the spheres in the aggregate \citep{Borghese94}.

In order to calculate the T-matrix of the whole aggregate, we define $\bmat{M}$ as the matrix of the coefficients defined by Eq.~(\ref{eq_deltas}). Then, the formal solution of Eq.~(\ref{eq_lin_system}) is:
\begin{equation}
  \mathcal{A}^{(p)}_{\eta \alpha l m} = -\sum_{p' l' m'} \left[ \bmat{M}^{-1} \right]^{(p p')}_{\alpha l m \alpha' l' m'} \mathcal{W}^{(p')}_{\eta \alpha' l' m'}. \label{eq_form_sol}
\end{equation}
Moreover, the addition theorem for vector multipole fields also allows us to write the scattered field of Eq.~(\ref{eq_sca_exp}) in terms of multipole fields with origin coincident with the origin of the coordinate system $O$:
\begin{equation}
  \bvec{E}_{S\eta} = E_{0 \eta} \sum_{p l m} \sum_{\alpha'} \sum_{p' l' m'} \bvec{H}^{(p)}_{lm}(\bvec{r}, k) \mathcal{J}^{(p p')}_{0 l m \alpha' l' m'} \mathcal{A}^{(p')}_{\eta \alpha' l' m'},
\end{equation}
though this solution only applies out of the smallest sphere that contains the whole particle (i.e., it is appropriate to calculate the scattered field in the far zone). As a consequence the amplitudes of the field scattered by the whole aggregate take the form:
\begin{equation}
  A^{(p)}_{\eta' l m} = \sum_{\alpha'} \sum_{p' l' m'} \mathcal{J}^{(p p')}_{0 l m \alpha' l' m'} \mathcal{A}^{(p')}_{\eta' \alpha' l' m'}, \label{eq_alt_ampl}
\end{equation}
where, this time, the quantities $\mathcal{J}^{(p p')}_{0 l m \alpha' l' m'}$ are the elements of the matrix that transfers the origin of the $\bvec{H}$-multipole fields from $\bvec{R}_{\alpha'}$ back to the origin of the coordinates. At this point, comparing Eq.~(\ref{eq_intro_tm}) with Eq.~(\ref{eq_alt_ampl}), using Eq.~(\ref{eq_form_sol}) to express $\mathcal{A}^{(p')}_{\eta' \alpha' l' m'}$ and Eq.~(\ref{eq_shifted_ampl}) to express $\mathcal{W}^{(p)}_{\eta \alpha l m}$, we obtain the T-matrix elements of the whole aggregate as:
\begin{equation}
  S^{(p p')}_{l m l' m'} = \sum_{\alpha \alpha'} \sum_{q L M} \sum_{q' L' M'} \mathcal{J}^{(p q)}_{0 l m \alpha L M} \left[ \bmat{M}^{-1} \right]^{q q'}_{\alpha L M \alpha' L' M'} \mathcal{J}^{(q' p')}_{\alpha' L' M' 0 l' m'}.
\end{equation}

The T-matrix elements depend on the reference frame used to describe the scattering process. However, since the vector spherical multipoles are eigenfunctions of the rotation matrices, their orientational averages can be obtained analytically. Knowledge of the T-matrix allows for the calculation of the scattering amplitude of Eq.~(\ref{eq_norm_ampl}) for any given combination of particle geometry, incident and scattered radiation field directions and, finally, to derive the scattering cross-section of an arbitrary particle through Eq.~(\ref{eq_diff_cs}). Therefore, the T-matrix acts as an analytical operator that can be used to calculate both differential and integrated cross-sections \citep{Saija03b, Borghese07}.

\subsection{Dynamical effects}
The processes of scattering, absorption and emission of radiation by interstellar dust affect the dynamical and thermal properties of the grains. In particular, scattering and absorption, which result in the global effect of extinction, depend critically on the cross-section of particles, as seen by the incident radiation field. The interaction of a radiation field with intensity $I_0$, propagating along a direction of incidence $\hat{\bvec{k}}_I$ on a non-spherical particle leads to a net mechanical force \citep{Mishchenko01, Polimeno18}:
\begin{equation}
  \bvec{F} = \frac{I_0}{c} \left[\sigma_{EXT} \hat{\bvec{k}}_I - \int_\Omega \frac{\de \sigma_{SCA}}{\de \Omega} \hat{\bvec{k}}_I \de \Omega \right].
\end{equation}
In general, $\bvec{F}$ is not oriented along the same direction of incidence $\hat{\bvec{k}}_I$, as it depends also on the scattering efficiency. Its component along the $\hat{\bvec{k}}_I$ direction is the radiation pressure force. For a scattering angle $\Phi$, such component can be expressed as:
\begin{equation}
  F_{pr} = \bvec{F} \cdot \hat{\bvec{k}}_I = \frac{I_0}{c} \left[ \sigma_{EXT} - \int_\Omega \cos \Phi \frac{\de \sigma_{SCA}}{\de \Omega} \de \Omega \right] = \frac{I_0}{c} [ \sigma_{EXT} - g \sigma_{SCA} ], \label{eq_rad_pres}
\end{equation}
where we introduced the asymmetry parameter for scattering at an angle $\Phi$:
\begin{equation}
  g = \frac{1}{\sigma_{SCA}} \int_\Omega \cos \Phi \frac{\de \sigma_{SCA}}{\de \Omega} \de \Omega.
\end{equation}
If the distribution of particle geometry and orientations can be represented in terms of analytical models, the T-matrix formalism allows for the calculation of the average expected cross-sections, leading to re-write the net radiation pressure effect of Eq.~(\ref{eq_rad_pres}) as:
\begin{equation}
  \langle F_{pr} \rangle = \frac{I_0}{c} [ \langle \sigma_{EXT} \rangle - \langle g \rangle \langle \sigma_{SCA} \rangle ], \label{eq_pres_mean}
\end{equation}
where the terms in brackets represent direction averaged quantities \citep{Polimeno21}.

\begin{figure}
  \begin{center}
    \includegraphics[width=0.8\textwidth]{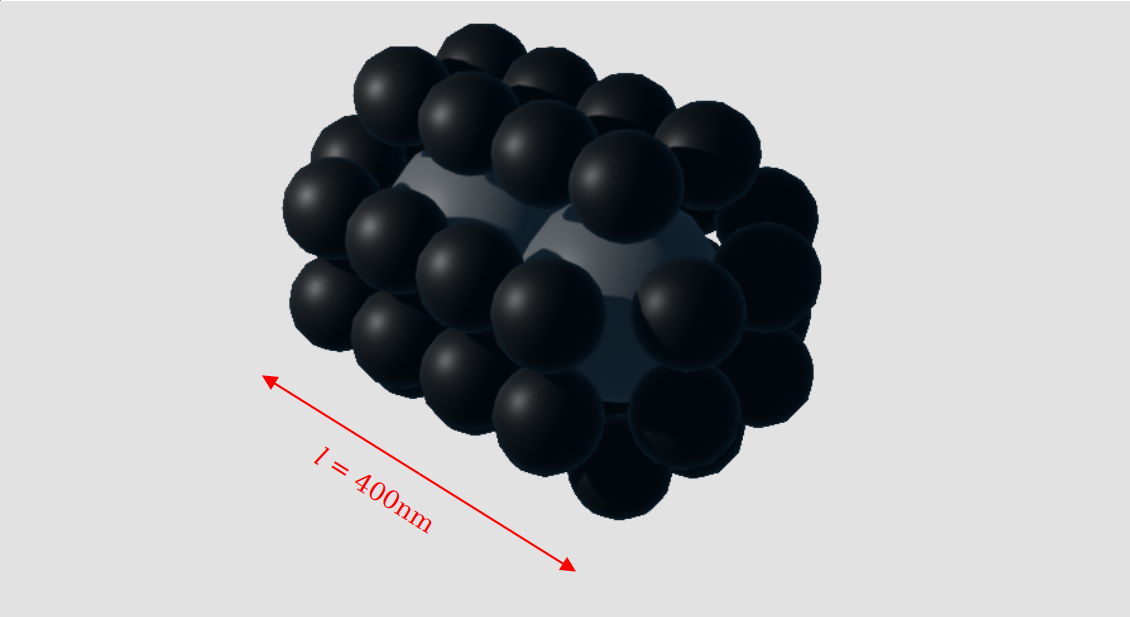}
  \end{center}
  \caption{Model particle used in this study. The particle is composed by a core structure, made by $2$ spheres, each with a radius $\rho_{core} = 68.63\,$nm, represented as grey-shaded spheres, and a mantle composed of $40$ spheres, each with a radius of $\rho_{mantle} = 0.5 \rho_{core}$ (black spheres). The size of the particle along its largest extension is $l = 400\,$nm. \label{fig_particle}}
\end{figure}
\section{Model implementation}\label{sec_model}

From the discussion presented in the previous section and, in particular, from Eq.~(\ref{eq_form_sol}), it turns out that the solution of the scattering problem in arbitrary geometry requires the inversion of the matrix $\bmat{M}$. This matrix can be large, since its dimensions depend on the number of spheres that are used to represent the aggregate and on the value of the maximum order where the calculation of the multipole field expansion - which is in principle an infinite series - is truncated. The value of the expansion order can be chosen on the basis of the size parameter, using numeric criteria that estimate the highest order required to obtain precision up to a predetermined number of significant digits \citep[e.g.][]{Wiscombe80}.

\begin{figure}
  \begin{center}
    \includegraphics[width=0.9\textwidth]{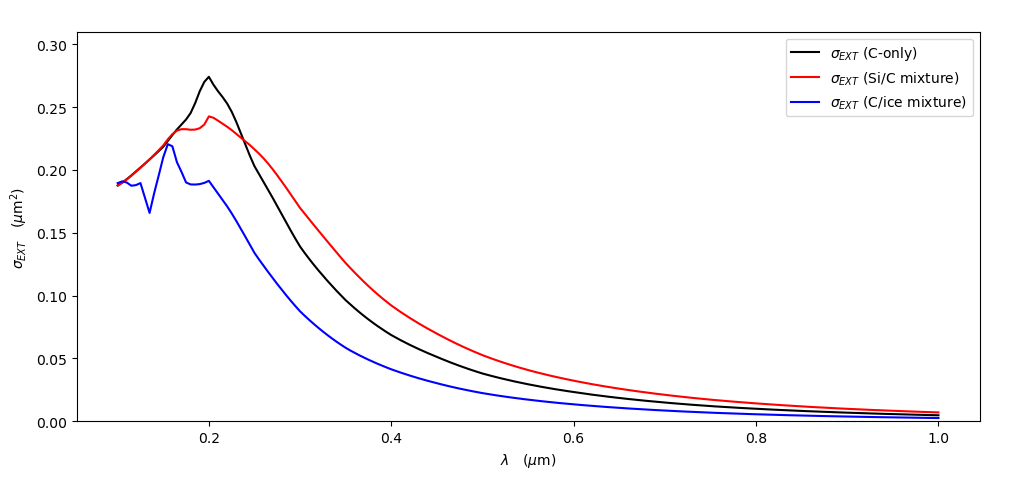}
  \end{center}
  \caption{Extinction cross-section $\sigma_{EXT}$ as a function of radiation wavelength $\lambda$ for a purely carbonaceous particle (black curve), a mixed particle composed by a silicate core with carbonaceous mantle (red curve) and a carbonaceous core with water ice coating (blue curve). \label{fig_models}}
\end{figure}
\glm{In this paper we aim at exploring the effects of different particle compositions and structural properties, with respect to the assumption of pure spherical symmetry, for models of interstellar dust grains computed within a parallel implementation of the T-matrix formalism.} \mai{The ideal model should be able to describe a sub-$\mu$m size particle possibly characterized by a layered and aggregated structure.  While this situation can in principle be met using a large number of small spherical units, we limit our first calculations to a test case that can still be easily handled with the use of commercial hardware. The assumed model of dust as clusters of a limited number of spherical monomers is physically motivated by coagulation processes in the diffuse interstellar medium, where grains grow through collisions that lead to loosely bound aggregates. Such structures are consistent with observed polarization trends and extinction curves, especially the correlation between $R_{V}$ and the wavelength of maximum polarization \citep{Wurm02}. Moreover, they naturally enhance extinction and radiation pressure cross sections per unit mass, making them efficient and physically realistic candidates for interstellar dust \citep{Iati04}.}

\glm{We therefore explore the extinction properties of an elongated particle, with a major axis length of $l = 400\,$nm, like the one shown in Fig.~\ref{fig_particle}, focusing on the wavelength range $100\, \mathrm{nm} \leq \lambda \leq 1\, \mu\mathrm{m}$.} The model represents a layered particle formed by an inner core, made up by $2$ spheres, each with a radius $\rho_{core} = 68.63\,$nm, and covered by a mantle made by $40$ spheres, each with a radius $\rho_{mantle} = 34.31\,$nm ($\rho_{mantle} = 0.5 \rho_{core}$). \glm{The multipolar field expansion needed to solve the particle as a whole should be computed up to a maximum order of $l_{max} = 15$. The $\bmat{M}$ matrix that needs to be computed and inverted to describe $42$ spherical elements up to $l_{max} = 15$, is} a $[21420 \times 21420]$ elements matrix that, for a double precision complex number representation, requires approximately $6.84\,$Gb of host memory to be stored. However, since the spherical monomers that compose the particle are smaller than the particle itself, instead of solving the problem all the way up to the maximum order, it is possible to adopt a lower internal truncation order $l_I$, to deal with the interactions among the field and the spherical monomers, while keeping a high external expansion order $l_E$, to treat the particle as a whole \citep{Saija03a, Iati04}. Following this approach, the size of the T-matrix is controlled by $l_I$, so that, using $l_I = 6$ as internal truncation order and $l_E = 15$ for the external one, $\bmat{M}$ reduces to a $[4032 \times 4032]$ elements object, which requires only $0.25\,$Gb of memory and can be more easily computed. We therefore derived the scattering, absorption and extinction cross-sections of our model particle, assuming three different cases: (i) a silicate core covered by a carbonaceous mantle; (ii) a carbonaceous core with a coating of water ice; (iii) a purely carbonaceous particle. We modeled the optical properties of the particles using the material permittivity given by \citet{Palik91} for carbonaceous materials, \citet{Draine84} for astronomical silicates and \citet{Warren08} for water ice. We compared the results of all three cases with the corresponding spherically symmetric cases, introducing an equivalent spherical particle defined as a layered sphere made up by the same amount in mass of the materials assumed for the three models.

\section{Results and discussion}\label{sec_results}
\begin{figure}
  \begin{center}
    \includegraphics[width=0.99\textwidth]{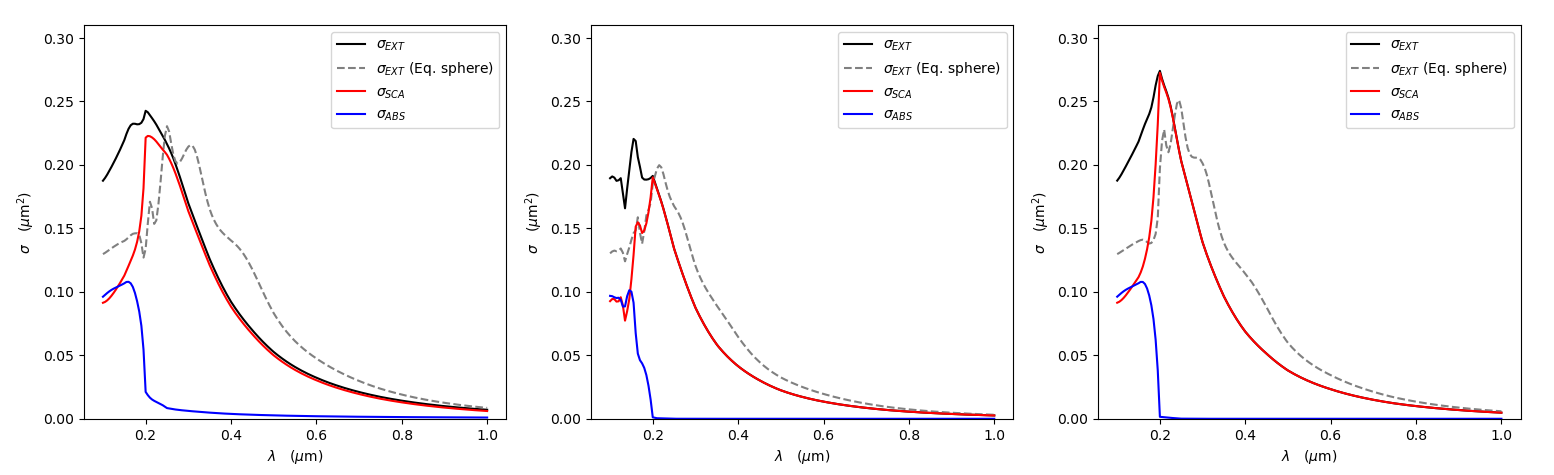}
  \end{center}
  \caption{Total extinction cross-sections $\sigma_{EXT}$ (black curves), scattering cross-sections $\sigma_{SCA}$ (red curves), and absorption cross-sections $\sigma_{ABS}$ (blue curves) for a silicate core / carbon mantle particle (left panel), a carbon core with ice coating (middle panel) and a purely carbonaceous particle (right panel). All models are compared with the corresponding equivalent spherical particles (grey dashed curves). \label{fig_comparison}}
\end{figure}

In Fig.~\ref{fig_models} we show the extinction cross-sections $\sigma_{EXT}$ obtained, as a function of wavelength, for randomly oriented particle distributions, using the model described in section~\ref{sec_model} \glm{for three different particle compositions}. The comparison of each particle model with the corresponding equivalent sphere, instead, is illustrated in the three panels of Fig.~\ref{fig_comparison}.

\begin{figure}
  \begin{center}
    \includegraphics[width=0.99\textwidth]{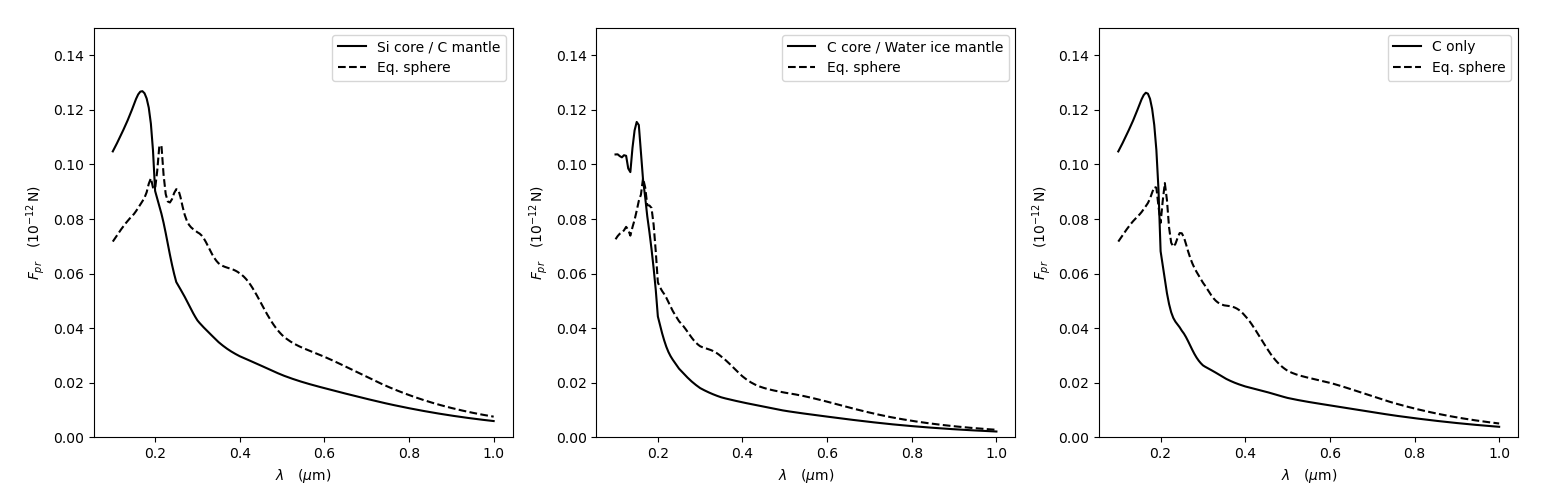}
  \end{center}
  \caption{Average radiation pressure force exerted on the model particle illustrated in Fig.~\ref{fig_particle} as a function of wavelength (continuous lines) compared with the pressure force expected for a spherical particle of the same mass (dashed lines) for the cases of a silicate core covered by a carbon mantle (left panel), a carbon particle with water ice coating (middle panel) and a purely carbonaceous particle (right panel). \label{fig_rad_pres}}
\end{figure}

\glm{While the test cases that were explored in this work are still constrained by the requirement to be solved with limited hardware resources, and therefore have to be considered preliminary, we can already observe that all the considered models are consistent with well established properties of interstellar extinction curves. In particular, all models exhibit a UV bump and they all converge to the Rayleigh regime in the long wavelength domain.} As illustrated in Fig.~\ref{fig_models}, the presence of layered structure and \glm{the introduction of} changes in composition can affect the overall shape of the cross-section, with icy-coated materials giving raise to a slightly lower overall extinction. The most interesting effects, however, are those resulting from the comparison of the cluster aggregates with the corresponding equivalent spheres, depicted in Fig.~\ref{fig_comparison}. In all cases, indeed, we observe dramatic differences in the cross-sections expected in the UV range. These substantial deviations of the extinction cross-sections from the predictions of spherically symmetric models also imply substantial differences in terms of dynamic effects, as it is shown by the comparison of the average radiation pressure force exerted on the particles, with respect to the spherically symmetric cases, depicted in Fig.~\ref{fig_rad_pres}. \glm{Here, we observe that the non-spherical particle models tend to experience a substantially stronger radiation pressure effect than the corresponding equivalent spheres in the short wavelength domain (specifically for $\lambda \leq 0.2\, \mu$m), while the situation reverts to the opposite at longer wavelengths. This finding agrees with the results reported by \citet{Saija03b}, who found a similar trend for particle models characterized by increasing deviations from spherical symmetry.}

Qualitatively, the observed result can be easily interpreted upon considering that short wavelength radiation probes irregularities at smaller scales than the long wavelength one. As a consequence, while all the tested combinations look quasi-spherical for long wavelength radiation, the presence of surface roughness, interstitial cavities and layered materials has a relevant impact on the UV domain. In presence of irregular structures, with dimensions matching the wavelength of the incoming radiation, strong scattering occurs. In our model this happens in the UV range where the incoming radiation becomes more subject to multiple scattering events across the particle structure. In addition, due to the substantial absorbing properties of materials in this wavelength range, we observe that particle models characterized by increasing structural complexity are subject to substantially different energy exchange rates with radiation fields, implying relevant effects on the survivability of volatile components like ice coatings. It is therefore clear that the assumption of over-simplified particle models may lead to incorrect interpretation of the role played by dust in particular spectral ranges.

\section{Conclusions}\label{sec_conclusions}
The interaction of interstellar dust with the surrounding environment is likely a key factor that needs to be accounted for, in order to properly understand the physics of ISM. Although there is wide agreement on the idea that interstellar dust is mainly in the form of sub-$\mu$m grains of mixed silicate and carbonaceous materials, more advanced models are required to explain detailed observations that show differences and evolution of the various dust components. Here \glm{we describe the first results obtained by applying a new parallel implementation of the T-matrix formalism, designed to reduce the calculation time by adapting to scalable computing hardware, to a set of test models. The considered test cases aim at evaluating the effects of changes in particle structure and composition on the predicted cross-sections. The considered models are} still quite simple and the results are, therefore, preliminary. It is nonetheless possible to draw the following conclusions:
\begin{description}
\item{i)} particle structural properties have a large impact on interaction with radiation having $\lambda \sim \rho$;
\item{ii)} evaluation of the cross-section of UV radiation with sub-$\mu$m grains needs accurate particle models;
\item{iii)} short wavelength radiation is extremely sensitive to properties such as \glm{small scale surface structure and material layering}.
\end{description}

\glm{The results presented in this work have been computed using the \texttt{NP\_TMcode} implementation of the T-matrix method.\footnote{\url{https://www.ict.inaf.it/gitlab/giacomo.mulas/np\_tmcode}}. The advantage of this code is that it provides an application of the T-matrix formalism over parallel architectures, which can be easily scaled from personal workstation to large computing facilities, allowing for the solution of increasingly complex models.} Furthermore, this method can be extended to describe the mechanical effects of light on dust particles and their dynamics \citep{Polimeno21, Magazzu23}. \glm{The possibility to run models on parallel architectures greatly reduces the amount of time needed to compute the properties of potentially large sets of scattering particles, characterized by a wide variety of structures and chemical compositions, allowing for a systematic application of the T-matrix formalism to more detailed investigations of the properties of interstellar dust.}

\section*{Acknowledgments}

Supported by Italian Research Center on High Performance Computing Big Data and Quantum Computing (ICSC), project funded by European Union - NextGenerationEU - and National Recovery and Resilience Plan (NRRP) - Mission 4 Component 2 within the activities of Spoke 3 (Astrophysics and Cosmos Observations) and by the Italian Minstry of University and Research (MUR), through PRIN Exo-CASH (PRIN 2022 project no. 2022J7ZFRA). This work was completed in part at the CINECA GPU HACKATHON 2024, part of the Open Hackathons program. The authors would like to acknowledge OpenACC-Standard.org for their support. The \texttt{NP\_TMcode} software is the parallel implementation of the T-matrix formalism, currently under development at INAF - Astronomical Observatory of Cagliari, on the basis of the codes originally written by Borghese, Denti and Saija. The authors gratefully acknowledge the journal reviewers for providing comments and suggestions that improved the text.

\bibliographystyle{jasr-model5-names}
\biboptions{authoryear}
\bibliography{ms_v03}

\end{document}